\documentclass{elsarticle}
\usepackage{lineno,hyperref,amsmath,amssymb,color,booktabs}

\journal{Physics Letters B}
\newcommand*\CentralInterval{\ensuremath{2.38\cdot 10^{-17} < \sigma_\text{rel} < 0.23}}
\newcommand*\UpperLimit{\ensuremath{\sigma_\text{rel}<0.20}}
\newcommand*\CentralIntervalConstrained{\ensuremath{10^{-14} \lesssim \sigma_\text{rel} < 0.23}}

\usepackage{xpatch}
\xpretocmd{\eqref}{Eq.~}{}{}
\newcommand*\patchAmsMathEnvironmentForLineno[1]{%
	\expandafter\let\csname old#1\expandafter\endcsname\csname #1\endcsname
	\expandafter\let\csname oldend#1\expandafter\endcsname\csname end#1\endcsname
	\renewenvironment{#1}%
	{\linenomath\csname old#1\endcsname}%
	{\csname oldend#1\endcsname\endlinenomath}}%
\newcommand*\patchBothAmsMathEnvironmentsForLineno[1]{%
	\patchAmsMathEnvironmentForLineno{#1}%
	\patchAmsMathEnvironmentForLineno{#1*}}%
\AtBeginDocument{%
	\patchBothAmsMathEnvironmentsForLineno{equation}%
	\patchBothAmsMathEnvironmentsForLineno{align}%
	\patchBothAmsMathEnvironmentsForLineno{flalign}%
	\patchBothAmsMathEnvironmentsForLineno{alignat}%
	\patchBothAmsMathEnvironmentsForLineno{gather}%
	\patchBothAmsMathEnvironmentsForLineno{multline}%
}

\begin{document}
\begin{frontmatter}
\title{Study of the wave packet treatment of neutrino oscillation at Daya Bay}
\def\ECUST{1}
\def\Wisconsin{2}
\def\Yale{3}
\def\BNL{4}
\def\NTU{5}
\def\NUU{6}
\def\NJU{7}
\def\IHEP{8}
\def\CUHK{9}
\def\NCTU{10}
\def\SDU{11}
\def\TsingHua{12}
\def\NCEPU{13}
\def\SZU{14}
\def\ZSU{15}
\def\Dubna{16}
\def\Siena{17}
\def\IIT{18}
\def\UIUC{19}
\def\LBNL{20}
\def\SJTU{21}
\def\BNU{22}
\fntext[BCC]{Now at: Department of Chemistry and Chemical Technology, Bronx Community College, Bronx, New York  10453, USA}
\def\UH{24}
\def\VirginiaTech{25}
\def\CIAE{26}
\def\USTC{27}
\def\NanKai{28}
\def\UC{29}
\def\DGUT{30}
\def\UCB{31}
\def\HKU{32}
\def\Charles{33}
\def\Princeton{34}
\def\XJTU{35}
\def\CUC{36}
\def\CalTech{37}
\def\WM{38}
\def\TempleUniversity{39}
\def\RPI{40}
\def\CGNPG{41}
\def\NUDT{42}
\def\IowaState{43}
\def\CQU{44}
\author[\ECUST]{F.~P.~An}
\author[\Wisconsin]{A.~B.~Balantekin}
\author[\Yale]{H.~R.~Band}
\author[\BNL]{M.~Bishai}
\author[\NTU,\NUU]{S.~Blyth}
\author[\NJU]{D.~Cao}
\author[\IHEP]{G.~F.~Cao}
\author[\IHEP]{J.~Cao}
\author[\IHEP]{W.~R.~Cen}
\author[\CUHK]{Y.~L.~Chan}
% ...

\date{\today}

\begin{abstract}
\noindent 
The disappearance of reactor $\bar{\nu}_e$ observed by the Daya Bay experiment is examined in the framework of a model in which the neutrino is described by a wave packet with a relative intrinsic momentum dispersion $\sigma_\text{rel}$. Three pairs of nuclear reactors and eight antineutrino detectors, each with good energy resolution, distributed among three experimental halls, supply a high-statistics sample of $\bar{\nu}_e$ acquired at nine different baselines.  This provides a unique platform to test the effects which arise from the wave packet treatment of neutrino oscillation. The modified survival probability formula was used to fit  Daya Bay  data, providing the first experimental limits:  \CentralInterval{}. 
Treating the dimensions of the reactor cores and detectors as constraints, the limits are improved: \CentralIntervalConstrained{}, and an upper limit of \UpperLimit{} is obtained. All limits correspond to  a 95\% C.L.
Furthermore, the effect due to the wave packet nature of neutrino oscillation is found to be insignificant for reactor antineutrinos detected by the Daya Bay experiment thus ensuring an unbiased measurement of the oscillation parameters $\sin^22\theta_{13}$ and $\Delta m^2_{32}$ within the plane wave model.
\end{abstract}		
\end{frontmatter}
	
\section{Introduction}
\subsection{Neutrino oscillation in the plane wave approximation}
The neutrino, a light electrically neutral fermion participating in weak interactions, was suggested by Pauli to save the conservation of energy and momentum in nuclear $\beta$-decays. Since then, three flavors of neutrinos  $\nu_\alpha = (\nu_e, \nu_\mu, \nu_\tau)$ were discovered, each produced or detected in association with a corresponding lepton $\ell_\alpha=(e,\mu,\tau)$. The neutrinos, which are completely parity-violating  in their weak interactions, suggested that the gauge group of the electro-weak sector of the remarkably successful Standard Model (SM) should be built using fermions with left-handed chirality. 
Given the unique properties of neutrinos, studies of them may reveal a path to physics beyond the SM. In the past, experiments observing solar and atmospheric neutrinos brought increased attention to neutrino physics due to long-standing discrepancies between detection rates and no-oscillation models. Despite an impressive number of proposed solutions to these problems, all were successfully resolved by the  hypothesis of neutrino oscillation, first proposed by Pontecorvo~\cite{Pontecorvo:1957cp,Pontecorvo:1957qd} in the late 1950's. Neutrino oscillation is a phenomenon firmly established in experiment, which has been observed with solar~\cite{Cleveland:1998nv,Kaether:2010ag,Abdurashitov:2009tn}, atmospheric~\cite{Fukuda:1998mi,Adamson:2014vgd}, particle accelerator~\cite{Ahn:2002up,Adamson:2014vgd} and reactor~\cite{Abe:2008aa,An:2012eh,RENO,Abe:2012tg} neutrinos.

Neutrino oscillation is a quantum phenomenon of quasi-periodic change of neutrino flavor $\nu_\alpha\to\nu_\beta$ with time. This phenomenon originates in the non-equivalence of neutrino flavor $\nu_\alpha$ and mass  $\nu_k=(\nu_1,\nu_2,\nu_3)$ eigenstates, differences in their masses, and an assumption that the produced and detected neutrino states are coherent superpositions of neutrino mass eigenstates:
\begin{equation}
|\nu_\alpha(p)\rangle=\sum_{k=1}^3 V_{\alpha k}^*|\nu_k(p)\rangle,
\label{eq:flavor_state_pw}
\end{equation}
where $V_{\alpha k}$ is an element of the unitary PMNS-matrix, named after Pontecorvo, Maki, Nakagawa, Sakata, and $p$ is the momentum of the neutrino. The time evolution of the state in~\eqref{eq:flavor_state_pw} is expressed as
\begin{equation}
|\nu_\alpha(t;p)\rangle=\sum_{k=1}^3 V_{\alpha k}^*\text{e}^{-iE_kt}|\nu_k(p)\rangle,
\label{eq:flavor_state_pw_time_evolution}
\end{equation} 
where $E_k=\sqrt{p^2+m_k^2}$. This leads to the oscillatory behavior of the probability to detect a neutrino originally of flavor $\alpha$ as having flavor $\beta$: 
\begin{equation}
P_{\alpha\beta}(L) = |\langle\nu_\beta(p)|\nu_\alpha(t;p)\rangle|^2= \sum_{k,j=1}^{3}V_{\alpha k}^* V_{\beta j}^* V_{\beta k}^{\phantom{*}} V_{\alpha j}^{\phantom{*}} \text{e}^{-iL/L^\text{osc}_{kj}},
\label{eq:planewave_prob}
\end{equation}
where $L^\text{osc}_{kj}= 4\pi p/\Delta m_{kj}^2$ is the oscillation length due to the non-zero differences $\Delta m^2_{kj}=m^2_k-m^2_j$, and time $t$ is approximated by the traveled distance $L$. 

The underlying theory, assuming a plane wave approximation, was developed in the middle of the 1970s~\cite{Eliezer:1975ja,Fritzsch:1975rz,Bilenky:1976yj}. Although successful in explaining a wide range of neutrino experiments, it is well known that this approximation is not self-consistent, and leads to a number of paradoxes~\cite{Akhmedov:2009rb,Giunti:2003ax}. The applicability of the plane wave approximation is discussed in detail in Refs.~\cite{Beuthe:2001rc,Akhmedov:2009rb,Giunti:2007ry,Bernardini:2004sw}. After the first theory was developed, Refs.~\cite{Nussinov:1976uw,Kayser:1981ye,Kiers:1995zj,Akhmedov:2012uu} pointed out the necessity of a wave packet treatment of neutrino oscillation. 

\subsection{Wave packet treatment of neutrino oscillation}
The wave packet is a coherent superposition of different waves whose momenta are distributed around the most probable value, with a certain ``width'' or dispersion. Therefore, a wave packet is localized in space-time as well as in energy-momentum space. The wave packet formalism facilitates the resolution of the paradoxes of the plane wave theory, and predicts the existence of a coherence length. The latter arises due to the different group velocities of a pair $\nu_k$ and $\nu_j$, which causes a separation in space over time. 
The smallness of the differences of neutrino masses relative to their typical energies suggests that the coherence length of neutrino oscillation is the largest among all known phenomena.


After the pioneering studies~\cite{Nussinov:1976uw,Kayser:1981ye,Kiers:1995zj}, the wave packet models of neutrino oscillation were developed in roughly two varieties. The  first one relies on a relativistic quantum mechanical (QM) formalism that does not predict the dispersion  of the neutrino wave packet in  momentum space, such as in Refs.~\cite{Beuthe:2001rc,Giunti:2007ry,Kayser:2010pr}. The second one is based on calculations within quantum field theory (QFT), describing all external particles involved in  neutrino production and detection  as wave packets while treating neutrinos  as virtual particles. The neutrino wave-function is  then calculated rather than postulated. The effective momentum dispersion of the neutrino wave function depends on the kinematics of neutrino production and detection and on the momentum dispersions of the external particles, as in Refs.~\cite{Grimus:1996av,Cardall:1999bz,Stockinger:2000sk,Beuthe:2002ej,Giunti:1993se,Akhmedov:2010ms,Naumov:2010um}. Both approaches predict a number of observable effects, like a quantitative condition on the coherence of mass eigenstates in the production-detection processes, as well as a loss of  coherence. 

In wave packet models, the intrinsic momentum dispersion $\sigma_p$ of the neutrino wave packet is an effective quantity comprising the microscopic momenta dispersions of all particles involved in the production and detection of the neutrino. A non-zero value of $\sigma_p$ leads with time to the {\em  decoherence} in the quantum superposition of massive neutrinos which results in a vanishing oscillation pattern of $\nu_\alpha\to\nu_\beta$ transitions. In addition, the oscillation pattern is smeared further in the reconstructed energy spectrum due to a non-zero experimental resolution $\delta_E$ of the neutrino energy.

Despite considerable progress in building wave packet models, none of these approaches provides a solid quantitative theoretical estimate of $\sigma_p$ or of the spatial width $\sigma_x=1/2\sigma_p$. Theoretical estimates vary by orders of magnitude, associating the dispersion of the neutrino wave packet with various scales; for example, uranium nucleus size  ($\sigma_x \simeq 10^{-11}$ cm, $\sigma_p\simeq 1$ MeV), atomic or inter-atomic size ($\sigma_x \simeq (10^{-8}-10^{-7})$ cm, $\sigma_p\simeq (10^3-10^2)$ eV), pressure broadening ($\sigma_x \simeq 10^{-4}$ cm, $\sigma_p\simeq 0.1$ eV), etc. 
While the current literature does not include calculations of the neutrino wave function from first principles for any type of neutrino experiment~\footnote{Recently, a first calculation which consistently treats  the full pion-neutrino-environment quantum system and calculates the decoherence effects for neutrinos produced in two-body decays was published in Ref.~\cite{Jones:2014sfa}}, it also lacks experimental investigations of  decoherence effects in neutrino oscillation inferred from the finite size of the neutrino wave function~\footnote{Attention to the decoherence phenomena in neutrino oscillation is increasing and the literature discusses possible decoherence effects due to physics beyond the SM like quantum gravity~\cite{Lisi:2000zt,Araki:2004mb,Barenboim:2006xt,Adamson:2008zt}, differing from the considerations of this paper, which studies the  consequences of a self-consistent way to describe neutrino oscillation within the Standard Model.}.

One of the motivations of this paper is to provide a first search for a possible loss of coherence in the quantum state of neutrinos following from the wave packet treatment of neutrino oscillations, using data from the Daya Bay Reactor Neutrino Experiment. The second motivation is to demonstrate that the oscillation parameters estimated with the plane wave approximation are unbiased.
The oscillation probability formula modified by the wave packet contribution, which is discussed further, has two distinctive features: it depends on $\Delta m^2_{kj}/p^2\sigma_\text{rel}$ via the so-called localization term and on $L\Delta m^2_{kj}\sigma_\text{rel}/p$ via the term responsible for the loss of coherence with  distance. The large statistics, good energy resolution, and multiple baselines of the Daya Bay experiment make its data valuable in the study of these quantum decoherence effects in neutrino oscillation.
	
\section{Analysis}
\subsection{Neutrino oscillation in a wave packet model}
\label{sec:wp_model}
Measured energy spectra of $\bar{\nu}_e$ interactions are compared to a prediction using a QM wave packet model of neutrino oscillation which is briefly outlined in what follows. We simplify the consideration by examining a one-dimensional wave packet of the neutrino~\footnote{While a neutrino travels in the three-dimensional space, the transverse part of its wave function essentially leads to the to $1/L^2$ dependence of the flux~\cite{Naumov:2013vea} and does not affect significantly the oscillation pattern.}. The plane wave state in \eqref{eq:flavor_state_pw} is replaced by a wave packet  describing a neutrino produced as flavor $\alpha$:
\begin{equation}
|\widetilde{\nu}_\alpha(p_P;t_P,x_P)\rangle = \sum_{k=1}^3 V_{\alpha k}^*\int\frac{dp}{2\pi} f_P(p) \text{e}^{-i\phi_{P}(p)}| \nu_k(p) \rangle,
\label{eq:wavepacket_1d}
\end{equation}
with $\phi_{P}(p) = E_k t_P - p x_P$. $f_P(p)$ is the  wave function of the neutrino in  momentum space and is assumed to be Gaussian: 
\begin{equation}
f_P(p) = \left (\frac{2\pi}{\sigma_{pP}^2}\right)^{\frac{1}{4}} \text{e}^{-\frac{(p-p_P)^2}{4\sigma^2_{pP}}},
\label{eq:gaussian_momentum_1d}
\end{equation}
where the subscript $P$ in $f_P(p)$, $p_P$ and $\sigma_{pP}$ indicates the quantities at production. In  configuration space the state in \eqref{eq:wavepacket_1d} describes a wave packet with mean coordinate $x_P$ at time $t_P$. The state in \eqref{eq:wavepacket_1d} is normalized as $\langle\widetilde{\nu}_\alpha(p_P;t_P,x_P)|\widetilde{\nu}_\alpha(p_P;t_P,x_P)\rangle=1$.  Similarly, a wave packet state at detection $|\widetilde{\nu}_\beta(p_D; t_D,x_D)\rangle$ is defined as the state given by~\eqref{eq:wavepacket_1d}.

A projection of $|\widetilde{\nu}_\alpha(p_P;t_P,x_P)\rangle$  onto $\langle\widetilde{\nu}_\beta(p_D;t_D,x_D)|$  produces the flavor-changing amplitude 
\begin{equation}
\mathcal{A}_{\alpha\beta}(p;t_D-t_P, L,\sigma_p)\equiv \langle\widetilde{\nu}_\beta(p_D;t_D,x_D)|\widetilde{\nu}_\alpha(p_P;t_P,x_P)\rangle,
\label{eq:amplitude}
\end{equation} 
which depends on $L\equiv x_D-x_P$, time difference~$t_D-t_P$ and on the effective mean neutrino momentum $p$ and momentum dispersion $\sigma_p$ comprising the details of production and detection~\footnote{The momentum integral in~\eqref{eq:amplitude} is calculated by expanding $E_k=\sqrt{p^2+m_k^2}$ in a Taylor series up to  second order around the effective momentum given by~\eqref{eq:mean_momentum_sigma}.} 
\begin{equation}
p = \frac{p_P \sigma_{pD}^2 + p_D  \sigma_{pP}^2}{\sigma_{pP}^2+\sigma_{pD}^2},\quad\frac{1}{\sigma_p^2}= \frac{1}{\sigma_{pP}^2}+\frac{1}{\sigma_{pD}^2}.
\label{eq:mean_momentum_sigma}
\end{equation}
The probability $|\mathcal{A}_{\alpha\beta}(p;t_D-t_P, L,\sigma_p)|^2$ should be integrated over production time $t_P$ (or, equivalently, over $t_D-t_P$) and most probable momentum $p_P$ to get an experimentally observable oscillation probability:
\begin{equation}
P_{\alpha\beta}(L)  =\sum_{k,\,j=1}^3\frac{ V^*_{\alpha k} V^{\phantom *}_{\beta k}V^{\phantom\dagger}_{ \alpha j}  V^*_{\beta j} }
{\sqrt[4]{1 +\left(L/L^{\text{d}}_{kj}\right)^2}}
\text{e}^{- \frac{\left(L/L^\text{coh}_{kj}\right)^2}{1+\left(L/L^{\text{d}}_{kj}\right)^2} -\mathrm{D}^2_{kj}}
\text{e}^{-i\widetilde{\varphi}_{kj}},
\label{eq:ossc}
\end{equation}
where the phase $\widetilde{\varphi}_{kj}$ is the sum of the plane wave phase $\varphi_{kj} = 2\pi L/L^\text{osc}_{kj}$ and correction $\varphi^d_{kj}$ due to the dispersion of the wave packet: $\widetilde{\varphi}_{kj} = \varphi_{kj} + \varphi^d_{kj}$, with 
\begin{equation}
\varphi^\text{d}_{kj} =
- \frac {L/L^\text{d}_{kj}}{1+\left(L/{L^\text{d}_{kj}}\right)^2}
\left(\frac L {L^\text{coh}_{kj}}\right)^2
+ \frac{1}{2} \arctan { \frac{L}{L^\text{d}_{kj}}}.
\label{eq:varphi_d}
\end{equation}
Oscillation probability formulas similar to~\eqref{eq:ossc} but neglecting  wave packet dispersion were obtained in several studies (see, for example, Refs.~\cite{Beuthe:2002ej,Beuthe:2001rc,Akhmedov:2010ms,Bernardini:2006ak}).
~\eqref{eq:ossc} has appeared as a particular case of a more general consideration within QFT with relativistic wave 
packets~\cite{Naumov:2010um}. Relativistic invariance suggests that $\sigma_\text{rel}$ should be Lorentz invariant. In the QM approach adopted in \eqref{eq:wavepacket_1d}-\eqref{eq:ossc}  the only possibility to preserve  Lorentz invariance is for $\sigma_\text{rel}$ to be a constant 
\footnote{Since the QFT approach considers both neutrino production and detection one finds that $\sigma_\text{rel}$, being a relativistic invariant, is actually a function of kinematic variables
involved in the production and detection processes as well as of momentum dispersions of wave packets describing all involved particles~\cite{Naumov:2013bea}.  Therefore, in comparing the QM and QFT approaches, we may treat the QM \protect{$\sigma_\text{rel}$} as that of the QFT approach averaged over the kinematic variables of all external wave packets involved in neutrino production and detection.}.
The probability in \eqref{eq:ossc} contains three quantities with dimensions of length:
\begin{equation}
\begin{aligned}
L^\text{osc}_{kj} & = \frac{4\pi p}{\Delta m^2_{kj}}, \qquad
L^\text{coh}_{kj} & =\frac{L^\text{osc}_{kl}}{\sqrt 2 \pi\sigma_\text{rel}}, \qquad L^\text{d}_{kj} & = \frac{L^\text{coh}_{kj}}{2\sqrt{2}\sigma_{\text{rel}}},
\end{aligned}
\label{lengthts-vacuum_1}
\end{equation}
where $\sigma_\text{rel}=\sigma_p/p$, $L^\text{osc}_{kj}$ is the usual oscillation length of a pair of neutrino states $|\nu_k\rangle$ and $|\nu_j\rangle$, $L^\text{coh}_{kj}$ is interpreted as the neutrino coherence length, i.e. the distance at which the interference of neutrino mass eigenstates vanishes, and finally $L^\text{d}_{kj}$ is the  dispersion length, i.e. a distance at which the wave packet is doubled in its spatial dimension due to the dispersion of waves moving with different velocities. The term 
\begin{equation}
\mathrm{D}^2_{kj} = \frac{1}{2} \left( \frac{\Delta m^2_{kj}}{4 p^2\sigma_\text{rel}} \right)^2=  \frac{1}{4} 
\left( \frac{\Delta m^2_{kj}}{\sigma_{m^2}} \right)^2 =\left( \frac{\sqrt{2}\pi\sigma_x}{L_{kj}^\text{osc}} \right)^2
\label{eq:D_factor}
\end{equation}
suppresses the coherence of massive neutrino states $|\nu_k\rangle$ and $|\nu_j\rangle$ if $\Delta m^2_{kj}\gg\sigma_{m^2}$, where $\sigma_{m^2}= 2\sqrt{2}p\sigma_p$ could be interpreted as an uncertainty in the neutrino mass squared~\cite{Kayser:1981ye}. 
$\mathrm{D}^2_{kj}$ can be seen from another perspective as the localization term
suppressing the oscillation if $\sqrt{2}\pi\sigma_x\gg L_{kj}^\text{osc}$, where $\sigma_x=(2\sigma_p)^{-1}$ is the width of neutrino wave packet in the configuration space. 

It is worth mentioning that terms in~\eqref{eq:ossc} which correspond to the interference of $\nu_k$ and $\nu_j$ states also get suppressed by the denominator  $\sqrt[4]{1 +\left(L/L^{\text{d}}_{kj}\right)^2}$ and vanish for both limits $\sigma_p\to 0$ and $\sigma_p\to\infty$, reducing the oscillation probability in~\eqref{eq:ossc} to the non-coherent sum
\begin{equation}
P_{\alpha\beta}=\sum_k|V_{\alpha k}|^2|V_{\beta k}|^2,
\label{eq:prob_decoherent}
\end{equation} 
which does not depend on energy and distance.

For the $\bar{\nu}_e$ at Daya Bay, $1-P_{\rm ee}$ is expressed as
\begin{equation}
\begin{aligned}
& \phantom {+}\cos^2\theta_{12}\sin^2\theta_{12}\cos^4\theta_{13}
\biggl(1-\frac {\exp{\left[-\frac{\left(L /L^\text{coh}_{21}\right)^2}{1+\left(L/{L^\text{d}_{21}}\right)^2}-D^2_{21}\right]}}
{\sqrt[4]{1 + \left(L/ {L^\text{d}_{21}}\right)^2}}\cos{(\varphi_{21}}+\varphi_{21}^\text{d})\biggr)\\
& +\cos^2\theta_{12}\cos^2\theta_{13}\sin^2\theta_{13}
\biggl(1-\frac {\exp{\left[-\frac{\left(L /L^\text{coh}_{31}\right)^2}{1+\left(L/{L^\text{d}_{31}}\right)^2}-D^2_{31}\right]}}
{\sqrt[4]{1 + \left(L/ {L^\text{d}_{31}}\right)^2}}\cos{(\varphi_{31}}+\varphi_{31}^\text{d})\biggr)\\
& +\sin^2\theta_{12}\cos^2\theta_{13}\sin^2\theta_{13}
\biggl(1-\frac {\exp{\left[-\frac{\left(L /L^\text{coh}_{32}\right)^2}{1+\left(L/{L^\text{d}_{32}}\right)^2}-D^2_{32}\right]}}
{\sqrt[4]{1 + \left(L/ {L^\text{d}_{32}}\right)^2}}\cos{(\varphi_{32}}+\varphi_{32}^\text{d})\biggr).
\end{aligned}
\label{eq:pee_wp}
\end{equation}

\subsection{Sensitivity of Daya Bay experiment to neutrino wave packet}
The Daya Bay experiment is composed of two near underground experimental halls (EH1 and EH2) and one far underground hall (EH3). Each of the experimental halls hosts identically designed antineutrino detectors (ADs). EH1 and EH2 contain two ADs each, while EH3 contains four ADs. Electron antineutrinos are produced in three pairs of nuclear reactors via $\beta$ decays of neutron-rich daughters of the fission isotopes ${}^{235}\text{U}$, ${}^{238}\text{U}$, ${}^{239}\text{Pu}$ and ${}^{241}\text{Pu}$, and detected via the inverse $\beta$ decay (IBD). The coincidence of the prompt ($e^+$ ionization and annihilation) and delayed ($n$ capture on Gd) signals efficiently suppresses the backgrounds, which amounted to less than 2\% (5\%) of the IBD candidates in the near (far) halls~\cite{An:2015rpe}. The Gd-doped liquid scintillator target is a cylinder of three meters in both height and diameter. The detectors have a light yield of about 165 photoelectrons/MeV and a reconstructed energy resolution  $\delta_E/E\approx 8\%$ at 1 MeV of deposited energy in the scintillator. More details on the experimental setup are contained in Refs.~\cite{An:2015rpe, DayaBay:2012aa,Dayabay:2014vka,An:2015qga}.

The studies in this paper are based on data acquired in the 6-AD period when there were two ADs in EH1, one AD in EH2 and 3 ADs in EH3, with the addition of the 8-AD period from October 2012 to November 2013, a total of 621 days. The number of IBD candidates used in this analysis, and the mean baselines of the three experimental halls to each pair of reactor cores, are summarized in Table~\ref{tab:data}.
\begin{table}[!htbp]
	\setlength{\tabcolsep}{6pt}
	\centering
	\begin{tabular}{ccrrr}
		\toprule
		& & \multicolumn{3}{c}{Mean distance, m} \\
		\cmidrule{3-5}
		Halls & IBD candidates & Daya Bay & Ling Ao & Ling Ao II \\
		\midrule
		EH1 & 613813 & 365 & 860 & 1310 \\
		EH2 & 477144 & 1348 & 481 & 529 \\
		EH3 & 150255 & 1909 & 1537 & 1542 \\
		\bottomrule
	\end{tabular}
	\caption{The number of IBD candidates and mean distances of the three experimental halls to the pairs of reactor cores.}
	\label{tab:data}
\end{table}
The expected numbers of IBD events are convolutions of the reactor-to-target expectation  with the detector-response function. The reactor-to-target  expectation takes into account the antineutrino fluxes from each reactor core including non-equilibrium and spent nuclear fuel corrections, first order in $1/m_p$ ($m_p$=proton mass) IBD cross-section accounting for the  positron emission angle~\cite{Vogel:1999zy}, and the oscillation survival probability $P_\text{ee}$ given by~\eqref{eq:planewave_prob} for the plane wave model and by~\eqref{eq:ossc} for the wave packet model. The detector response-function accounts for energy loss in the inner acrylic vessel, liquid scintillator and electronics non-linearity and energy resolution $\delta_E$.

For relatively large values of $\sigma_p\simeq\delta_E$, the effects of these two parameters on the observed energy spectra might appear similar, however they are distinct. First, they have different physical origins: while $\sigma_p$ is governed by the most localized particle in the production and detection of the neutrino, $\delta_E$ is determined by  the energy depositions of the final state particles in the detector. Second, these effects can also be distinguished from their order of occurrence since the microscopic processes used in the energy estimation occur later in time with respect to the neutrino interaction in the detector. Third, their effects are not identical. In particular, as described in Sec.~\ref{sec:wp_model}, the limit $\sigma_p\to 0$ leads to the decoherence of neutrino oscillation in contrast to the impact of energy resolution which does not lead to any smearing in the reconstructed energy spectrum in the limit $\delta_E\to 0$.

In order to illustrate analytically an interplay of $\sigma_p$ and $\delta_E$, let us consider the exponential in the oscillation probability in~\eqref{eq:ossc} convolved with a Gaussian energy resolution, as a function of the reconstructed energy $E_{\rm vis}$, assuming $\delta_E\ll p$, infinite dispersion length $L^\text{d}$, neglecting the $D^2$ term,  and suppressing mass eigenstate indices for the sake of compactness~\footnote{The actual implementation of the detector effects in this analysis was performed numerically without approximations}:
\begin{equation}
\begin{aligned}
&\frac{1}{\sqrt{2\pi}\delta_E}\int dp \; \exp{\left(-i \; 2\pi L/L^\text{osc} - \left(L/L^\text{coh}\right)^2 - (p-E_{\rm vis})^2/2\delta_E^2\right)} \\
&\simeq \exp{\left(-i \; 2\pi L/L^\text{osc}_\text{rec} - \left(L/L^\text{coh}_\text{eff}\right)^2\right)},
\end{aligned}
\end{equation}
where $L^\text{osc}$ and $L^\text{coh}$ are given by~\eqref{lengthts-vacuum_1} and the effective coherence length comprises both the intrinsic $\sigma_p$ and detector resolution $\delta_E$:
\begin{eqnarray}
\left(\frac{1}{L^\text{coh}_\text{eff}}\right)^2 = \left(\frac{1}{L^\text{coh}_\text{rec}}\right)^2 + \left(\frac{1}{L^\text{coh}_\text{det}}\right)^2,
\end{eqnarray}
where $L^\text{osc}_\text{rec}$ and $L^\text{coh}_\text{rec}$ are given by $L^\text{osc}$ and $L^\text{coh}$ replacing $p$ with $E_{\rm vis}$, and $L^\text{coh}_\text{det}$ is given by $L^\text{coh}_\text{rec}$, replacing $\sigma_p$ with $\delta_E$. The interplay of $\sigma_p$ and $\delta_E$ is illustrated by the effective coherence length $L^\text{coh}_\text{eff}$, which is 
dominantly determined by the smallest among $L^\text{coh}_\text{rec}$ and $L^\text{coh}_\text{det}$, or by the largest among $\sigma_p$ and $\delta_E$.

The following provides simple numerical estimates of wave packet effects on neutrino oscillations at Daya Bay. For a typical  momentum of $p=4$ MeV of detected reactor $\bar{\nu}_e$, the oscillation would be suppressed for two distinctive domains of $\sigma_\text{rel}$. The domain $\sigma_\text{rel}\gtrsim O(0.1)$ corresponds to significant contributions from $L$--dependent interference-suppressing terms and corrections to the oscillation phase $\varphi^d_{32}$ in \eqref{eq:ossc}, while the $D^2_{kj}$ term is negligibly small. For example, at $L=L_{32}^\text{osc}/2$ the exponential suppression reaches  its maximum $\text{e}^{-\pi/8}$ at $\sigma_\text{rel}=1/\sqrt{2\pi}\simeq 0.4$. Correspondingly, the coherence and dispersion lengths read $L_{32}^\text{coh}\simeq 2.2$ km and $L_{32}^\text{d}\simeq 2$ km. At larger values of $\sigma_\text{rel}$ and at a fixed distance the spatial dispersion of neutrino wave packets partially compensates the loss of coherence due to the spatial separation of $\nu_k$ and $\nu_j$.  

The domain $\sigma_\text{rel}\lesssim  O(2.8\cdot 10^{-17})$ corresponds to $D_{32}^2 \gtrsim 1$, which is significant in suppressing the interference in \eqref{eq:ossc} through the $L$--independent term, while the $L$--dependent terms are negligibly small. Thus, the region of $O(2.8\cdot 10^{-17}) \ll \sigma_\text{rel} \ll O(0.1)$ is where the wave packet impact on neutrino oscillation is negligible for the Daya Bay experiment. 

\begin{figure}[htb]
	\begin{center}
		\includegraphics[width=0.9\textwidth]{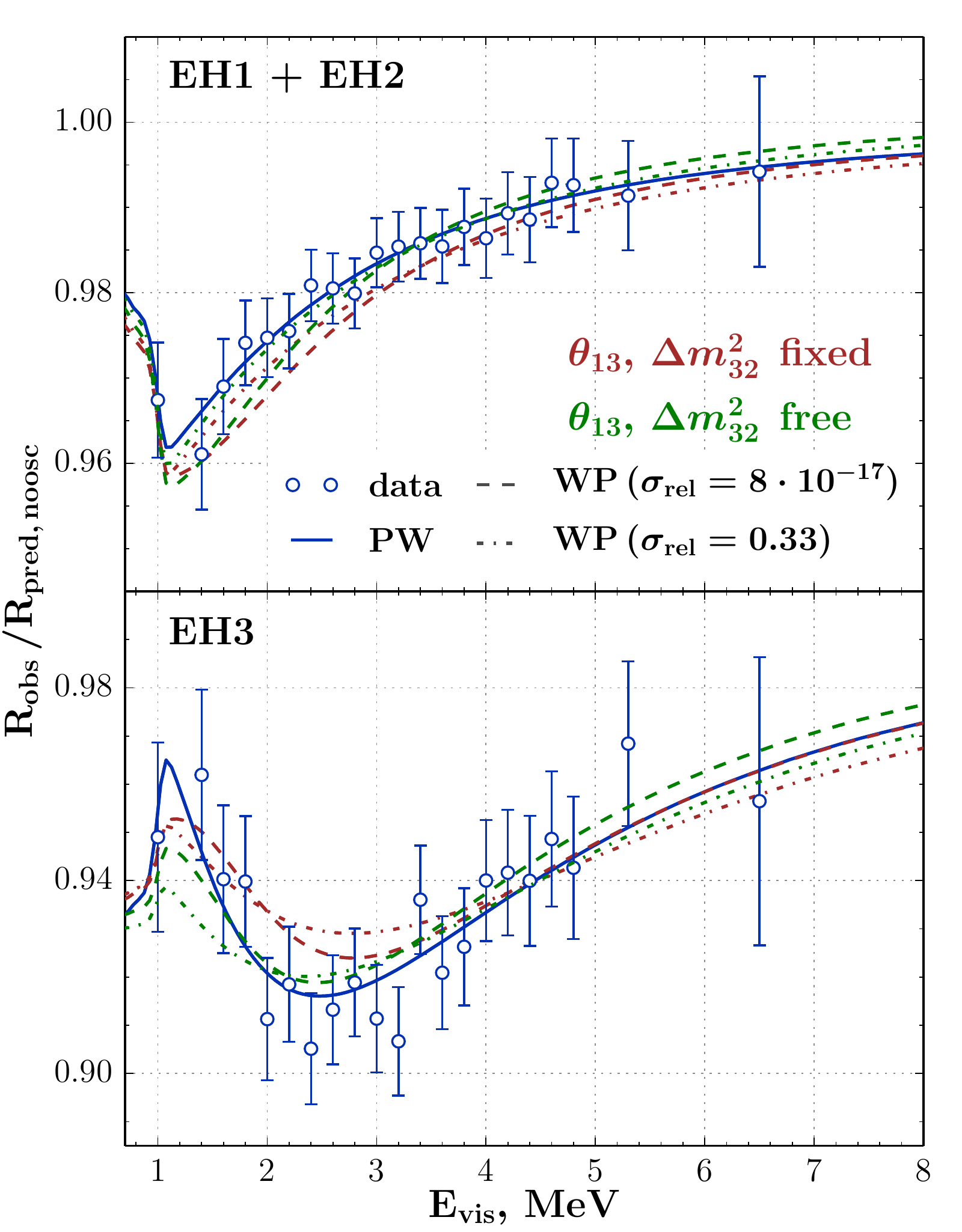}
		\caption{\label{fig:Data2Theory}
			Ratios of the observed to expected numbers of IBD events in the absence of oscillation  as a function of reconstructed visible energy $E_\text{vis}$.  The data are grouped by near (EH1+EH2) and far (EH3) halls, displayed in the upper and in the bottom panels respectively, with the error bars representing the statistical uncertainties. Superimposed solid lines are ratios assuming neutrino oscillations within the plane wave model (PW) with the best-fit values of $\sin^22\theta_{13}$ and $\Delta m^2_{32}$ obtained with the plane wave model. 
			The ratios using the wave-packet model (WP) assume $\sigma_\text{rel} = 0.33$ (dashed line) and  $\sigma_\text{rel}= 8\cdot 10^{-17}$ (dot-dashed line), as two examples. 
			The green lines correspond to the wave packet model ratios assuming the best-fit values of $\sin^22\theta_{13}$ and $\Delta m^2_{32}$ obtained with the plane wave model and thus, inconsistent with the data by about five standard deviations.  The red lines correspond to the wave packet model ratios assuming the best-fit values of $\sin^22\theta_{13}$ and $\Delta m^2_{32}$ obtained within the wave packet model, yielding a much better agreement with the data.  
			All ratios enter the region below $2m_e$, which corresponds to the IBD threshold, because of detector response effects like energy reconstruction and absorption in the inner acrylic vessel (see details in Refs.~\cite{An:2015rpe,An:2015qga}).
			}
	\end{center}	
\end{figure}

For illustrative purposes Fig.~\ref{fig:Data2Theory} shows the ratio of the observed to expected numbers of IBD events assuming no oscillation using the data collected at the near and far experimental halls as a function of reconstructed visible energy $E_\text{vis}$. Figure~\ref{fig:Data2Theory} also shows the expected ratio for neutrino oscillation with the plane wave and wave packet models with $\sigma_\text{rel}$ of $0.33$ and $8\cdot 10^{-17}$ as examples.

Both model expectations are shown with the oscillation parameters fixed to their best-fit values within the plane wave model~\footnote{The following values of the oscillation parameters were used in Fig.~\ref{fig:Data2Theory}: $\Delta m^2_{21} = 7.53 \cdot 10^{-5}\text{ eV}^2$, $\Delta m^2_{32}=2.45\cdot 10^{-3}\text{ eV}^2$, $\sin^22\theta_{12}=0.846$, $\sin^22\theta_{13}=0.0852$.}. For this set of parameters, the wave packet models with $\sigma_\text{rel}=0.33$ and with $\sigma_\text{rel}=8\cdot 10^{-17}$ are inconsistent with the data by about five standard deviations, thus motivating the chosen values of $\sigma_\text{rel}$. The two panels illustrate how the visible energy spectra are modified in the near and far halls depending on the intrinsic dispersion of the neutrino wave packet. Remarkably, most changes in the energy spectra due to $\sigma_\text{rel}$ are in opposite directions for near and far halls, which can be explained qualitatively as follows. As mentioned above, the extremes $\sigma_p\to 0$ and $\sigma_p\to\infty$ would yield fully decoherent neutrinos with the oscillation probability given by~\eqref{eq:prob_decoherent}. Antineutrinos detected at the near halls experience a relatively small oscillation in the plane wave approach. The values of $\sigma_\text{rel}$ selected for Fig.~\ref{fig:Data2Theory} make the $\bar{\nu}_e$ partially decoherent and $P_\text{ee}$ tend towards \eqref{eq:prob_decoherent}, predicting a {\em smaller} number of surviving $\bar{\nu}_e$ as compared to the plane wave formula. The distance at which the far detectors of the Daya Bay experiment are placed is tuned to observe the maximal oscillation effect due to $\Delta m^2_{32}$. Partial decoherence of the $\bar{\nu}_e$ tends to reduce the oscillation, thus predicting a {\em larger} number of survived $\bar{\nu}_e$ with respect to the plane wave formula.  This feature of Daya Bay provides additional sensitivity to the decoherence effects and makes such a study less sensitive to the predicted reactor $\bar{\nu}_e$ spectrum. 

The data can be reasonably well described by 
\begin{equation}
\begin{aligned}
\Delta m^2_{32}=2.17\cdot 10^{-3}\text{ eV}^2, \quad  \sin^22\theta_{13}=0.102,\\
\sigma_\text{rel}=8\cdot 10^{-17}, \quad \chi^2/\text{ndf} = 246.8/(256-4),
\end{aligned}
\label{eq:example_left}
\end{equation}
and by 
\begin{equation}
\begin{aligned}
\Delta m^2_{32}=2.16\cdot 10^{-3}\text{ eV}^2, \quad \sin^22\theta_{13}=0.097,\\
\sigma_\text{rel}=0.33, \quad        \chi^2/\text{ndf} = 253.8/(256-4).
\end{aligned}
\label{eq:example_right}
\end{equation}
These results demonstrate that one could obtain reasonable fits of the data within the wave packet model with certain values of $\sigma_\text{rel}$ and yield best-fit values of the oscillation parameters which differ from the corresponding best-fit values with the plane wave model, assuming normal mass hierarchy~\footnote{The best-fit values of the oscillation parameters $\sin^22\theta_{13}$ and $\Delta m^2_{32}$ are different from our previous publication~\cite{An:2015rpe} because of a  different implementation of systematic uncertainties and another choice of $E_\text{vis}$ binning.}:
\begin{equation}
\begin{aligned}
\Delta m^2_{32}=2.45\cdot 10^{-3}\text{ eV}^2,\quad \sin^22\theta_{13}=0.0852, \\
\phantom {\sigma_\text{rel}=0.33} \quad\chi^2/\text{ndf} = 245.9/(256-3). 
\end{aligned}
\label{eq:true_osc_minimum}
\end{equation}
However, Eqs.~\ref{eq:example_left},~\ref{eq:example_right} do not correspond to the global minimum of the $\chi^2$ discussed below because $\sigma_\text{rel}$  was fixed to two  arbitrary values for illustrative purposes. In order to find the global minimum we performed a detailed statistical analysis of the allowed region of $\sigma_\text{rel}$. 

\subsection{Statistical framework}
As the goodness-of-fit measure we use $\chi^2(\boldsymbol{\eta}) = (\mathbf{d}-\mathbf{t}(\boldsymbol{\eta}))^TV^{-1}(\mathbf{d}-\mathbf{t}(\boldsymbol{\eta}))$, where $\mathbf{d}$ is a data vector containing detected numbers of IBD candidates in  energy bins and in different detectors, while $\mathbf{t}(\boldsymbol{\eta})$  is the corresponding theoretical model vector which depends on constrained and unconstrained parameters $\boldsymbol{\eta}$. All constraints of the model as well as expected fluctuations in the number of IBD events  are encompassed in the covariance matrix $V$. The model vector $\mathbf{t}(\boldsymbol{\eta})$ comprises expected numbers of IBD and background events.  All constrained parameters (or systematic uncertainties) relevant for the Daya Bay oscillation analyses were taken into account in this analysis. These  are mainly associated with the reactor antineutrino flux, background predictions and the detector response modeling. The uncertainty of the detector response is dominant. Details can be found in Refs.~\cite{An:2015rpe,An:2015qga}.

The analysis was done with four unconstrained parameters $\sigma_{\text{rel}}$, $\Delta m^2_{32}$, $\sin^22\theta_{13}$ and reactor flux normalization $N$.  The confidence regions are produced by means of two statistical methods: the conventional fixed-level $\Delta \chi^2$ analysis and the Feldman-Cousins method~\cite{Feldman:1997qc}. The marginalized $\Delta \chi^2$ statistic is
\begin{equation}
\Delta \chi^2(\boldsymbol{\eta}') = \min\limits_{\boldsymbol{\eta}\setminus\boldsymbol{\eta}'} \chi^2(\boldsymbol{\eta}) -\min\limits_{\boldsymbol{\eta}}\chi^2(\boldsymbol{\eta}),
\label{eq:statistic}
\end{equation}
where $\boldsymbol{\eta}=(\sigma_{\text{rel}}, \Delta m^2_{32}, \sin^22\theta_{13}, N)$ and $\boldsymbol{\eta}'$ is its subspace with parameters of interest ($\boldsymbol{\eta}'=\sigma_\text{rel}$ for one dimensional interval, and $\boldsymbol{\eta}'=(\sigma_\text{rel},\Delta m^2_{32}$) or $\boldsymbol{\eta}'=(\sigma_\text{rel},\sin^22\theta_{13}$) for two dimensional regions), and both are used to determine the $p$-value of the observed dataset and the model. 

The closed interval corresponding to the $100\cdot(1-\alpha)\%$ confidence level (C.L.) is constructed for both the fixed-level $\Delta \chi^2$ analysis and the Feldman-Cousins method as the region of $\boldsymbol{\eta}'$ which satisfies:
\begin{equation}
\Delta\chi^2(\boldsymbol{\eta}') < \Delta\chi^2_{1-\alpha},
\label{eq:interval_construction}
\end{equation}
where $\Delta\chi^2_{1-\alpha}$ is the $(1-\alpha)$-th quantile of the statistic in~\eqref{eq:statistic}. The tabulated values of the quantile $\chi^2_{n;1-\alpha}$ of the  $\chi^2_n$ distribution with $n$ degrees of freedom ($n=1,2$ for one and two dimensional confidence regions) were used for the fixed-level $\Delta \chi^2$ analysis. Toy Monte Carlo sampling was used to determine $\Delta\chi^2_{1-\alpha}$ of the statistic in~\eqref{eq:statistic} with the Feldman-Cousins method.

An open confidence interval can be constructed if neutrinos are assumed to be produced and detected coherently, which is equivalent to assuming $\sigma_\text {rel} \gg 10^{-16}$. In this case, instead of using~\eqref{eq:statistic}, an upper bound on $\sigma_{rel}$ can be computed using the modified statistic
~\cite{Cowan:2010js}
\begin{equation}
\Delta\chi^2_\text{up}(\sigma_{\text{rel}}) = \begin{cases}
\Delta\chi^2(\sigma_{\text{rel}}) &\mbox{if } \hat{\sigma}_{\text{rel}} < \sigma_{\text{rel}} \\
0 &\mbox{if } \hat{\sigma}_{\text{rel}} > \sigma_{\text{rel}},
\end{cases}
\label{eq:1sided_statistic}
\end{equation}
with $\hat{\sigma}_{\text{rel}}$ representing the best-fit value.  In the fixed-level $\Delta \chi^2$ analysis the $100\cdot(1-\alpha)\%$ C.L. upper limit is given by:
\begin{equation}
\Delta\chi^2(\sigma_{\text{rel}}) \le \chi^2_{1;1-2\alpha}.
\end{equation}
For example, in order to set a 95\% C.L. upper limit, the quantile $\chi^2_{1;0.9}=2.71$ was used. The Feldman-Cousins method automatically produces the proper interval using the interval construction in~\eqref{eq:interval_construction}.

\section{Results and Discussion}
\begin{figure}[htb]
	\begin{center}
		\includegraphics[width=0.9\textwidth]{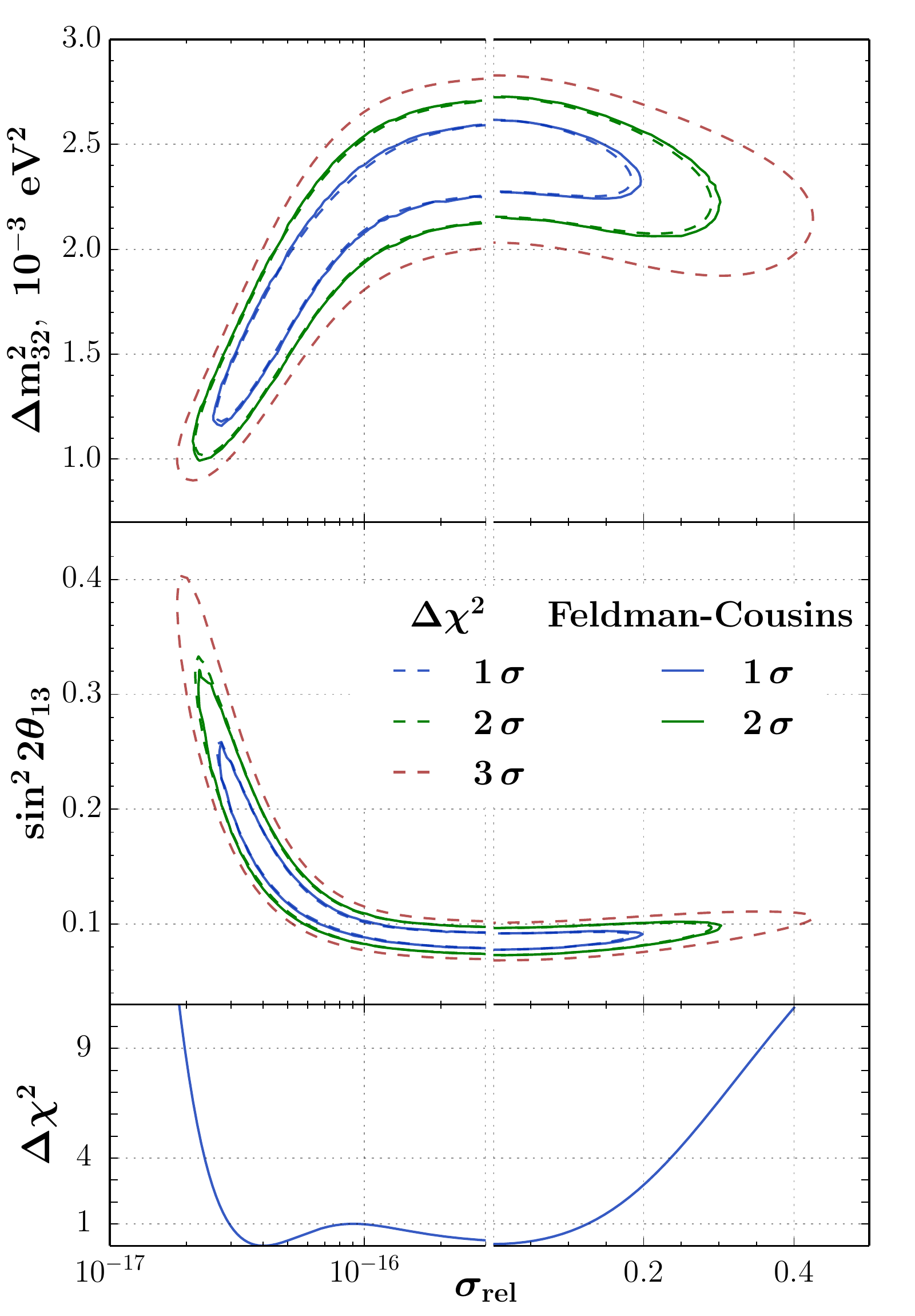}	    
		\caption{\label{fig:DeltaChi2_dm31_all}	Allowed regions of $(\Delta m^2_{32},\sigma_{\text{rel}})$ (top) and of $(\sin^2 2\theta_{13},\sigma_{\text{rel}})$ (middle) parameters obtained with fixed-level $\Delta\chi^2$ (contours corresponding to $1\sigma$, $2\sigma$, $3\sigma$ C.L., dashed lines) and within the Feldman-Cousins (contours corresponding to $1\sigma$, $2\sigma$ C.L., solid lines) methods. Bottom panel shows the marginalized $\Delta\chi^2(\sigma_\text{rel})$ statistic given by~\eqref{eq:statistic} vs $\sigma_\text{rel}$. Note the break in the abscissa and the change from a logarithmic to linear scale.}
	\end{center}
\end{figure}

Figure~\ref{fig:DeltaChi2_dm31_all} displays the allowed regions in $(\Delta m^2_{32},\sigma_{\text{rel}})$ and $(\sin^2 2\theta_{13},\sigma_{\text{rel}})$  obtained with both the fixed-level $\Delta\chi^2$ and the Feldman-Cousins methods, which are found to be consistent. For  the values of $\sigma_\text{rel}\lesssim 10^{-16}$ the decoherence effects lead to strong correlations between $\Delta m^2_{32}, \sin^22\theta_{13}$ and $\sigma_\text{rel}$, yielding smaller values of $\Delta m^2_{32}$ and larger values of $\sin^22\theta_{13}$. These correlations are expected taking into account the explicit form of $1-P_\text{ee}(L)$ in~\eqref{eq:pee_wp}. For $\sigma_\text{rel}\gtrsim O(0.1)$, these correlations are found to be significantly weaker. 

The best-fit point corresponds to
\begin{equation}
\begin{aligned}
\Delta m^2_{32}=1.59\cdot 10^{-3}\text{ eV}^2, \quad  \sin^22\theta_{13}=0.160,\\
\sigma_\text{rel}=4.0\cdot 10^{-17}, \quad \chi^2/\text{ndf} = 245.9/(256-4),
\end{aligned}
\label{eq:wp_fit_data}
\end{equation}
with the p-value $0.596$ which is smaller than the p-value $0.614$ with the plane wave model given by~\eqref{eq:true_osc_minimum}. The allowed region for $\sigma_\text{rel}$ at a 95\% C.L. reads:
\begin{equation}
\CentralInterval.
\label{eq:2sided_interval}
\end{equation}
The upper bound of~\eqref{eq:2sided_interval} corresponds to $L^\text {coh}_{32}>1.94\;L^\text {osc}_{32}/2$ and $L^\text {d}_{32}>2.96\;L^\text {osc}_{32}/2$. The lower bound can also be interpreted in terms of length $\sigma_x$ which corresponds to the spatial width of the neutrino wave packet. Taking the average momentum $p=4$ MeV of detected reactor $\bar{\nu}_e$, the lower bound of~\eqref{eq:2sided_interval} rules out $\sigma_x\gtrsim 1$ km. The Daya Bay data is not sensitive enough to constrain the $D^2_{kj}$ term significantly better.

Thus, the lower limit is much weaker than an obvious constraint of $\sigma_x\lesssim 2$ m which follows from the consideration that the $\sigma_x$ (which equals 1/2$\sigma_ p$) of $\bar{\nu}_e$ wave packets detected by the Daya Bay Experiment does not exceed the dimensions of the reactor cores and detectors. Taking this constraint into account, $\sigma_p\gtrsim 5\cdot 10^{-8}$ eV, which for the average momentum $p=4$ MeV, translates into $\sigma_\text{rel}\gtrsim 10^{-14}$. Such a $\sigma_\text{rel}$ corresponds to the regime where $D^2_{kj}\ll 1$ and the localization term can be safely neglected, thus allowing us to put an upper limit of:
\begin{equation}
\UpperLimit, \text{ at a } 95\%\text{ C.L.}
\end{equation}

\section*{Summary}
We performed a search for the footprint of the neutrino wave packet which should show itself through specific modifications of the  neutrino oscillation probability. The reported analysis of the Daya Bay data provides, for the first time, an allowed interval of the intrinsic relative dispersion of neutrino momentum \CentralInterval{}. Taking into account the actual dimensions of the reactor cores and detectors, we find that the lower limit $\sigma_{\rm rel} > 10^{-14}$  corresponds to the regime when the localization term is vanishing, thus allowing us to put an upper limit:
\UpperLimit{} at a 95\% C.L. 
The obtained limits can be read as $10^{-11}\text{ cm } \lesssim\sigma_x \lesssim 2$ m.

The current limits are dominated by statistics. With three years of additional data the upper limit on $\sigma_\text{rel}$ is expected to be improved by about 30\%.  
The allowed decoherence effect due to the wave packet nature of neutrino oscillation is found to be insignificant for reactor antineutrinos detected by the Daya Bay experiment thus ensuring an unbiased measurement of the oscillation parameters $\sin^22\theta_{13}$ and $\Delta m^2_{32}$ within the plane wave model.

\section*{Acknowledgements}
Daya Bay is supported in part by the Ministry of Science and Technology of China, the U.S. Department of Energy, the Chinese Academy of Sciences, the CAS Center for Excellence in Particle Physics, the National Natural Science Foundation of China, the Guangdong provincial government, the Shenzhen municipal government, the China General Nuclear Power Group, Key Laboratory of Particle and Radiation Imaging (Tsinghua University), the Ministry of Education, Key Laboratory of Particle Physics and Particle Irradiation (Shandong University), the Ministry of Education, Shanghai Laboratory for Particle Physics and Cosmology, the Research Grants Council of the Hong Kong Special Administrative Region of China, the University Development Fund of The University of Hong Kong, the MOE program for Research of Excellence at National Taiwan University, National Chiao-Tung University, and NSC fund support from Taiwan, the U.S. National Science Foundation, the Alfred~P.~Sloan Foundation, the Ministry of Education, Youth, and Sports of the Czech Republic, the Joint Institute of Nuclear Research in Dubna, Russia, the National Commission of Scientific and Technological Research of Chile, and the Tsinghua University
Initiative Scientific Research Program. We acknowledge Yellow River Engineering Consulting Co., Ltd., and China Railway 15th Bureau Group Co., Ltd., for building the underground laboratory. We are grateful
for the ongoing cooperation from the China General Nuclear Power Group and China Light and Power Company.

	

\begin{thebibliography}{10}
\expandafter\ifx\csname url\endcsname\relax
  \def\url#1{\texttt{#1}}\fi
\expandafter\ifx\csname urlprefix\endcsname\relax\def\urlprefix{URL }\fi
\expandafter\ifx\csname href\endcsname\relax
  \def\href#1#2{#2} \def\path#1{#1}\fi

\bibitem{Pontecorvo:1957cp}
B.~Pontecorvo, {Mesonium and anti-mesonium}, Sov. Phys. JETP 6 (1957) 429, [Zh.
  Eksp. Teor. Fiz.33,549(1957)].

\bibitem{Pontecorvo:1957qd}
B.~Pontecorvo, {Inverse beta processes and nonconservation of lepton charge},
  Sov. Phys. JETP 7 (1958) 172--173, [Zh. Eksp. Teor. Fiz.34,247(1957)].

\bibitem{Cleveland:1998nv}
B.~T. Cleveland, T.~Daily, R.~Davis, Jr., J.~R. Distel, K.~Lande, C.~K. Lee,
  P.~S. Wildenhain, J.~Ullman, {Measurement of the solar electron neutrino flux
  with the Homestake chlorine detector}, Astrophys. J. 496 (1998) 505--526.
\newblock \href {http://dx.doi.org/10.1086/305343} {\path{doi:10.1086/305343}}.

\bibitem{Kaether:2010ag}
F.~Kaether, W.~Hampel, G.~Heusser, J.~Kiko, T.~Kirsten, {Reanalysis of the
  GALLEX solar neutrino flux and source experiments}, Phys. Lett. B685 (2010)
  47--54.
\newblock \href {http://arxiv.org/abs/1001.2731} {\path{arXiv:1001.2731}},
  \href {http://dx.doi.org/10.1016/j.physletb.2010.01.030}
  {\path{doi:10.1016/j.physletb.2010.01.030}}.

\bibitem{Abdurashitov:2009tn}
J.~N. Abdurashitov, et~al., {Measurement of the solar neutrino capture rate
  with gallium metal. III: Results for the 2002--2007 data-taking period},
  Phys. Rev. C80 (2009) 015807.
\newblock \href {http://arxiv.org/abs/0901.2200} {\path{arXiv:0901.2200}},
  \href {http://dx.doi.org/10.1103/PhysRevC.80.015807}
  {\path{doi:10.1103/PhysRevC.80.015807}}.

\bibitem{Fukuda:1998mi}
Y.~Fukuda, et~al., {Evidence for oscillation of atmospheric neutrinos}, Phys.
  Rev. Lett. 81 (1998) 1562--1567.
\newblock \href {http://dx.doi.org/10.1103/PhysRevLett.81.1562}
  {\path{doi:10.1103/PhysRevLett.81.1562}}.

\bibitem{Adamson:2014vgd}
P.~Adamson, et~al., {Combined analysis of $\nu_{\mu}$ disappearance and
  $\nu_{\mu} \rightarrow \nu_{e}$ appearance in MINOS using accelerator and
  atmospheric neutrinos}, Phys. Rev. Lett. 112 (2014) 191801.
\newblock \href {http://dx.doi.org/10.1103/PhysRevLett.112.191801}
  {\path{doi:10.1103/PhysRevLett.112.191801}}.

\bibitem{Ahn:2002up}
M.~H. Ahn, et~al., {Indications of neutrino oscillation in a 250 km long
  baseline experiment}, Phys. Rev. Lett. 90 (2003) 041801.
\newblock \href {http://dx.doi.org/10.1103/PhysRevLett.90.041801}
  {\path{doi:10.1103/PhysRevLett.90.041801}}.

\bibitem{Abe:2008aa}
S.~Abe, et~al., {Precision Measurement of Neutrino Oscillation Parameters with
  KamLAND}, Phys. Rev. Lett. 100 (2008) 221803.
\newblock \href {http://arxiv.org/abs/0801.4589} {\path{arXiv:0801.4589}},
  \href {http://dx.doi.org/10.1103/PhysRevLett.100.221803}
  {\path{doi:10.1103/PhysRevLett.100.221803}}.

\bibitem{An:2012eh}
F.~An, et~al., {Observation of electron-antineutrino disappearance at Daya
  Bay}, Phys.Rev.Lett. 108 (2012) 171803.
\newblock \href {http://arxiv.org/abs/1203.1669} {\path{arXiv:1203.1669}},
  \href {http://dx.doi.org/10.1103/PhysRevLett.108.171803}
  {\path{doi:10.1103/PhysRevLett.108.171803}}.

\bibitem{RENO}
J.~Ahn, et~al., Phys. Rev. Lett. 108 (2012) 191802.

\bibitem{Abe:2012tg}
Y.~Abe, et~al., {Reactor electron antineutrino disappearance in the Double
  Chooz experiment}, Phys. Rev. D86 (2012) 052008.
\newblock \href {http://arxiv.org/abs/1207.6632} {\path{arXiv:1207.6632}},
  \href {http://dx.doi.org/10.1103/PhysRevD.86.052008}
  {\path{doi:10.1103/PhysRevD.86.052008}}.

\bibitem{Eliezer:1975ja}
S.~Eliezer, A.~R. Swift, {Experimental Consequences of electron
  Neutrino-Muon-neutrino Mixing in Neutrino Beams}, Nucl. Phys. B105 (1976)
  45--51.
\newblock \href {http://dx.doi.org/10.1016/0550-3213(76)90059-6}
  {\path{doi:10.1016/0550-3213(76)90059-6}}.

\bibitem{Fritzsch:1975rz}
H.~Fritzsch, P.~Minkowski, {Vector-Like Weak Currents, Massive Neutrinos, and
  Neutrino Beam Oscillations}, Phys. Lett. B62 (1976) 72--76.
\newblock \href {http://dx.doi.org/10.1016/0370-2693(76)90051-4}
  {\path{doi:10.1016/0370-2693(76)90051-4}}.

\bibitem{Bilenky:1976yj}
S.~M. Bilenky, B.~Pontecorvo, {Again on Neutrino Oscillations}, Lett. Nuovo
  Cim. 17 (1976) 569.
\newblock \href {http://dx.doi.org/10.1007/BF02746567}
  {\path{doi:10.1007/BF02746567}}.

\bibitem{Akhmedov:2009rb}
E.~K. Akhmedov, A.~Y. Smirnov, {Paradoxes of neutrino oscillations},
  Phys.Atom.Nucl. 72 (2009) 1363--1381.
\newblock \href {http://arxiv.org/abs/0905.1903} {\path{arXiv:0905.1903}},
  \href {http://dx.doi.org/10.1134/S1063778809080122}
  {\path{doi:10.1134/S1063778809080122}}.

\bibitem{Giunti:2003ax}
C.~Giunti, {Coherence and wave packets in neutrino oscillations},
  Found.Phys.Lett. 17 (2004) 103--124.
\newblock \href {http://arxiv.org/abs/hep-ph/0302026}
  {\path{arXiv:hep-ph/0302026}}, \href
  {http://dx.doi.org/10.1023/B:FOPL.0000019651.53280.31}
  {\path{doi:10.1023/B:FOPL.0000019651.53280.31}}.

\bibitem{Beuthe:2001rc}
M.~Beuthe, {Oscillations of neutrinos and mesons in quantum field theory},
  Phys.Rept. 375 (2003) 105--218.
\newblock \href {http://arxiv.org/abs/hep-ph/0109119}
  {\path{arXiv:hep-ph/0109119}}, \href
  {http://dx.doi.org/10.1016/S0370-1573(02)00538-0}
  {\path{doi:10.1016/S0370-1573(02)00538-0}}.

\bibitem{Giunti:2007ry}
C.~Giunti, C.~W. Kim, {Fundamentals of Neutrino Physics and Astrophysics},
  2007.

\bibitem{Bernardini:2004sw}
A.~E. Bernardini, S.~De~Leo, {An Analytic approach to the wave packet formalism
  in oscillation phenomena}, Phys.Rev. D70 (2004) 053010.
\newblock \href {http://arxiv.org/abs/hep-ph/0411134}
  {\path{arXiv:hep-ph/0411134}}, \href
  {http://dx.doi.org/10.1103/PhysRevD.70.053010}
  {\path{doi:10.1103/PhysRevD.70.053010}}.

\bibitem{Nussinov:1976uw}
S.~Nussinov, {Solar Neutrinos and Neutrino Mixing}, Phys. Lett. B63 (1976)
  201--203.
\newblock \href {http://dx.doi.org/10.1016/0370-2693(76)90648-1}
  {\path{doi:10.1016/0370-2693(76)90648-1}}.

\bibitem{Kayser:1981ye}
B.~Kayser, {On the Quantum Mechanics of Neutrino Oscillation}, Phys.Rev. D24
  (1981) 110.
\newblock \href {http://dx.doi.org/10.1103/PhysRevD.24.110}
  {\path{doi:10.1103/PhysRevD.24.110}}.

\bibitem{Kiers:1995zj}
K.~Kiers, S.~Nussinov, N.~Weiss, {Coherence effects in neutrino oscillations},
  Phys.Rev. D53 (1996) 537--547.
\newblock \href {http://arxiv.org/abs/hep-ph/9506271}
  {\path{arXiv:hep-ph/9506271}}, \href
  {http://dx.doi.org/10.1103/PhysRevD.53.537}
  {\path{doi:10.1103/PhysRevD.53.537}}.

\bibitem{Akhmedov:2012uu}
E.~Akhmedov, D.~Hernandez, A.~Smirnov, {Neutrino production coherence and
  oscillation experiments}, JHEP 1204 (2012) 052.
\newblock \href {http://arxiv.org/abs/1201.4128} {\path{arXiv:1201.4128}},
  \href {http://dx.doi.org/10.1007/JHEP04(2012)052}
  {\path{doi:10.1007/JHEP04(2012)052}}.

\bibitem{Kayser:2010pr}
B.~Kayser, J.~Kopp, {Testing the wave packet approach to neutrino oscillations
  in future experiments}\href {http://arxiv.org/abs/1005.4081}
  {\path{arXiv:1005.4081}}.

\bibitem{Grimus:1996av}
W.~Grimus, P.~Stockinger, {Real oscillations of virtual neutrinos}, Phys.Rev.
  D54 (1996) 3414--3419.
\newblock \href {http://arxiv.org/abs/hep-ph/9603430}
  {\path{arXiv:hep-ph/9603430}}, \href
  {http://dx.doi.org/10.1103/PhysRevD.54.3414}
  {\path{doi:10.1103/PhysRevD.54.3414}}.

\bibitem{Cardall:1999bz}
C.~Y. Cardall, D.~J. Chung, {The MSW effect in quantum field theory}, Phys.Rev.
  D60 (1999) 073012.
\newblock \href {http://arxiv.org/abs/hep-ph/9904291}
  {\path{arXiv:hep-ph/9904291}}, \href
  {http://dx.doi.org/10.1103/PhysRevD.60.073012}
  {\path{doi:10.1103/PhysRevD.60.073012}}.

\bibitem{Stockinger:2000sk}
P.~Stockinger, {Introduction to a field-theoretical treatment of neutrino
  oscillations}, Pramana 54 (2000) 203--214.
\newblock \href {http://dx.doi.org/10.1007/s12043-000-0017-1}
  {\path{doi:10.1007/s12043-000-0017-1}}.

\bibitem{Beuthe:2002ej}
M.~Beuthe, {Towards a unique formula for neutrino oscillations in vacuum},
  Phys.Rev. D66 (2002) 013003.
\newblock \href {http://arxiv.org/abs/hep-ph/0202068}
  {\path{arXiv:hep-ph/0202068}}, \href
  {http://dx.doi.org/10.1103/PhysRevD.66.013003}
  {\path{doi:10.1103/PhysRevD.66.013003}}.

\bibitem{Giunti:1993se}
C.~Giunti, C.~Kim, J.~Lee, U.~Lee, {On the treatment of neutrino oscillations
  without resort to weak eigenstates}, Phys.Rev. D48 (1993) 4310--4317.
\newblock \href {http://arxiv.org/abs/hep-ph/9305276}
  {\path{arXiv:hep-ph/9305276}}, \href
  {http://dx.doi.org/10.1103/PhysRevD.48.4310}
  {\path{doi:10.1103/PhysRevD.48.4310}}.

\bibitem{Akhmedov:2010ms}
E.~K. Akhmedov, J.~Kopp, {Neutrino oscillations: Quantum mechanics vs. quantum
  field theory}, JHEP 1004 (2010) 008.
\newblock \href {http://arxiv.org/abs/1001.4815} {\path{arXiv:1001.4815}},
  \href {http://dx.doi.org/10.1007/JHEP04(2010)008, 10.1007/JHEP10(2013)052}
  {\path{doi:10.1007/JHEP04(2010)008, 10.1007/JHEP10(2013)052}}.

\bibitem{Naumov:2010um}
D.~Naumov, V.~Naumov, {A Diagrammatic treatment of neutrino oscillations},
  J.Phys. G37 (2010) 105014.
\newblock \href {http://arxiv.org/abs/1008.0306} {\path{arXiv:1008.0306}},
  \href {http://dx.doi.org/10.1088/0954-3899/37/10/105014}
  {\path{doi:10.1088/0954-3899/37/10/105014}}.

\bibitem{Jones:2014sfa}
B.~Jones, {Dynamical pion collapse and the coherence of conventional neutrino
  beams}, Phys.Rev. D91~(5) (2015) 053002.
\newblock \href {http://arxiv.org/abs/1412.2264} {\path{arXiv:1412.2264}},
  \href {http://dx.doi.org/10.1103/PhysRevD.91.053002}
  {\path{doi:10.1103/PhysRevD.91.053002}}.

\bibitem{Lisi:2000zt}
E.~Lisi, A.~Marrone, D.~Montanino, {Probing possible decoherence effects in
  atmospheric neutrino oscillations}, Phys. Rev. Lett. 85 (2000) 1166--1169.
\newblock \href {http://arxiv.org/abs/hep-ph/0002053}
  {\path{arXiv:hep-ph/0002053}}, \href
  {http://dx.doi.org/10.1103/PhysRevLett.85.1166}
  {\path{doi:10.1103/PhysRevLett.85.1166}}.

\bibitem{Araki:2004mb}
T.~Araki, et~al., {Measurement of neutrino oscillation with KamLAND: Evidence
  of spectral distortion}, Phys. Rev. Lett. 94 (2005) 081801.
\newblock \href {http://arxiv.org/abs/hep-ex/0406035}
  {\path{arXiv:hep-ex/0406035}}, \href
  {http://dx.doi.org/10.1103/PhysRevLett.94.081801}
  {\path{doi:10.1103/PhysRevLett.94.081801}}.

\bibitem{Barenboim:2006xt}
G.~Barenboim, N.~E. Mavromatos, S.~Sarkar, A.~Waldron-Lauda, {Quantum
  decoherence and neutrino data}, Nucl. Phys. B758 (2006) 90--111.
\newblock \href {http://arxiv.org/abs/hep-ph/0603028}
  {\path{arXiv:hep-ph/0603028}}, \href
  {http://dx.doi.org/10.1016/j.nuclphysb.2006.09.012}
  {\path{doi:10.1016/j.nuclphysb.2006.09.012}}.

\bibitem{Adamson:2008zt}
P.~Adamson, et~al., {Measurement of Neutrino Oscillations with the MINOS
  Detectors in the NuMI Beam}, Phys.Rev.Lett. 101 (2008) 131802.
\newblock \href {http://arxiv.org/abs/0806.2237} {\path{arXiv:0806.2237}},
  \href {http://dx.doi.org/10.1103/PhysRevLett.101.131802}
  {\path{doi:10.1103/PhysRevLett.101.131802}}.

\bibitem{Naumov:2013vea}
D.~Naumov, {On the Theory of Wave Packets}, Phys.Part.Nucl.Lett. 10 (2013)
  642--650.
\newblock \href {http://arxiv.org/abs/1309.1717} {\path{arXiv:1309.1717}},
  \href {http://dx.doi.org/10.1134/S1547477113070145}
  {\path{doi:10.1134/S1547477113070145}}.

\bibitem{Bernardini:2006ak}
A.~E. Bernardini, M.~M. Guzzo, F.~R. Torres, {Second-order corrections to
  neutrino two-flavor oscillation parameters in the wave packet approach}, Eur.
  Phys. J. C48 (2006) 613.
\newblock \href {http://arxiv.org/abs/hep-ph/0612001}
  {\path{arXiv:hep-ph/0612001}}, \href
  {http://dx.doi.org/10.1140/epjc/s10052-006-0032-6}
  {\path{doi:10.1140/epjc/s10052-006-0032-6}}.

\bibitem{Naumov:2013bea}
V.~A. Naumov, D.~S. Shkirmanov, {Extended Grimus-Stockinger theorem and inverse
  square law violation in quantum field theory}, Eur. Phys. J. C73~(11) (2013)
  2627.
\newblock \href {http://arxiv.org/abs/1309.1011} {\path{arXiv:1309.1011}},
  \href {http://dx.doi.org/10.1140/epjc/s10052-013-2627-z}
  {\path{doi:10.1140/epjc/s10052-013-2627-z}}.

\bibitem{An:2015rpe}
F.~P. An, et~al., {New Measurement of Antineutrino Oscillation with the Full
  Detector Configuration at Daya Bay}, Phys. Rev. Lett. 115~(11) (2015) 111802.
\newblock \href {http://arxiv.org/abs/1505.03456} {\path{arXiv:1505.03456}},
  \href {http://dx.doi.org/10.1103/PhysRevLett.115.111802}
  {\path{doi:10.1103/PhysRevLett.115.111802}}.

\bibitem{DayaBay:2012aa}
F.~P. An, et~al., {A side-by-side comparison of Daya Bay antineutrino
  detectors}, Nucl. Instrum. Meth. A685 (2012) 78--97.
\newblock \href {http://arxiv.org/abs/1202.6181} {\path{arXiv:1202.6181}},
  \href {http://dx.doi.org/10.1016/j.nima.2012.05.030}
  {\path{doi:10.1016/j.nima.2012.05.030}}.

\bibitem{Dayabay:2014vka}
F.~P. An, et~al., {The muon system of the Daya Bay Reactor antineutrino
  experiment}, Nucl. Instr. Meth. A 773 (2015) 8--20.
\newblock \href {http://dx.doi.org/10.1016/j.nima.2014.09.070}
  {\path{doi:10.1016/j.nima.2014.09.070}}.

\bibitem{An:2015qga}
F.~P. An, et~al., {The Detector System of The Daya Bay Reactor Neutrino
  Experiment}, Nucl. Instrum. Meth. A811 (2016) 133--161.
\newblock \href {http://arxiv.org/abs/1508.03943} {\path{arXiv:1508.03943}},
  \href {http://dx.doi.org/10.1016/j.nima.2015.11.144}
  {\path{doi:10.1016/j.nima.2015.11.144}}.

\bibitem{Vogel:1999zy}
P.~Vogel, J.~F. Beacom, {Angular distribution of neutron inverse beta decay,
  $\bar{\nu}_e + p \to e^+ + n$}, Phys. Rev. D60 (1999) 053003.
\newblock \href {http://arxiv.org/abs/hep-ph/9903554}
  {\path{arXiv:hep-ph/9903554}}, \href
  {http://dx.doi.org/10.1103/PhysRevD.60.053003}
  {\path{doi:10.1103/PhysRevD.60.053003}}.

\bibitem{Feldman:1997qc}
G.~J. Feldman, R.~D. Cousins, {A Unified approach to the classical statistical
  analysis of small signals}, Phys.Rev. D57 (1998) 3873--3889.
\newblock \href {http://arxiv.org/abs/physics/9711021}
  {\path{arXiv:physics/9711021}}, \href
  {http://dx.doi.org/10.1103/PhysRevD.57.3873}
  {\path{doi:10.1103/PhysRevD.57.3873}}.

\bibitem{Cowan:2010js}
G.~Cowan, K.~Cranmer, E.~Gross, O.~Vitells, {Asymptotic formulae for
  likelihood-based tests of new physics}, Eur. Phys. J. C71 (2011) 1554,
  [Erratum: Eur. Phys. J.C73,2501(2013)].
\newblock \href {http://arxiv.org/abs/1007.1727} {\path{arXiv:1007.1727}},
  \href {http://dx.doi.org/10.1140/epjc/s10052-011-1554-0,
  10.1140/epjc/s10052-013-2501-z} {\path{doi:10.1140/epjc/s10052-011-1554-0,
  10.1140/epjc/s10052-013-2501-z}}.

\end{thebibliography}

%
%
	
\end{document}